\documentclass[10pt,conference]{IEEEtran}

\usepackage{research5}
\usepackage{psfrag}
\usepackage{epsfig}
\usepackage{graphicx}
\usepackage{multicol}

\allowdisplaybreaks[3]

\newtheorem{thm}{Theorem}
\newtheorem{lem}{Lemma}

\begin{document}

\title{Covering Point Patterns}

\author{\authorblockN{Amos Lapidoth}
\authorblockA{
ETH Zurich\\
8092 Zurich, Switzerland\\
\texttt{ amos.lapidoth@ethz.ch}}\and
\authorblockN{Andreas Mal\"ar}
\authorblockA{ETH Zurich\\ 8092 Zurich, Switzerland\\
\texttt{amalaer@ee.ethz.ch}}\and
\authorblockN{Ligong Wang}
\authorblockA{ETH Zurich\\ 8092 Zurich, Switzerland\\
\texttt{wang@isi.ee.ethz.ch}}}

\maketitle

\begin{abstract}
  An encoder observes a point pattern---a finite number of points in
  the interval $[0,T]$---which is to be described to a reconstructor
  using bits. Based on these bits, the reconstructor wishes to select
  a subset of $[0,T]$ that contains all the points in the pattern. It
  is shown that, if the point pattern is produced by a homogeneous
  Poisson process of intensity $\lambda$, and if the reconstructor is
  restricted to select a subset of average Lebesgue measure not
  exceeding $DT$, then, as $T$ tends to infinity, the minimum number
  of bits per second needed by the encoder is $-\lambda\log D$. It is
  also shown that, as $T$ tends to infinity, any point
  pattern on $[0,T]$ containing no more than $\lambda T$ points can be
  successfully described using $-\lambda \log D$ bits per second in
  this sense. Finally, a
  Wyner-Ziv version of this problem is considered where some of the
  points in the pattern are known to the reconstructor.
\end{abstract}

\section{Introduction}
An encoder observes a point pattern---a finite number of points in the
interval $[0,T]$---which is to be described to a reconstructor using
bits. Based on these bits, the reconstructor wishes to produce a
covering-set---a subset of $[0,T]$ containing all the points---of
least Lebesgue measure. There is a trade-off between the number of
bits used and the Lebesgue measure of the covering-set. This trade-off
can be formulated as a continuous-time rate-distortion problem
(Section~\ref{sec:poisson}). In this paper we investigate this
trade-off in the limit where $T\to\infty$.

When the point pattern is produced by a
homogeneous Poisson process, this problem is closely related to
that of transmitting information through an ideal peak-limited
Poisson channel 
\cite{kabanov78,davis80,wyner88,wyner88b}. In fact, the two problems
can be considered dual in the sense of
\cite{coverchiang02}. However, the duality results of
\cite{coverchiang02} only apply to discrete memoryless channels and
sources, so they cannot be directly used to solve our
problem. Instead, we shall use a technique that is similar to Wyner's
\cite{wyner88,wyner88b} to find the desired rate-distortion
function. We shall show that, if the point pattern is the 
outcome of a homogeneous Poisson process of intensity $\lambda$, and
if the reconstructor is restricted to select covering-sets of average
measure not exceeding $DT$, then the minimum number of
bits per second needed by the encoder to describe the pattern is
$-\lambda\log D$. 

Previous works \cite{rubin74,colemankiyavashsubramanian08} have
studied rate-distortion functions of the Poisson process with different
distortion measures. It is interesting to 
notice that our rate-distortion function, $-\lambda\log D$, is
equal to the one in \cite{colemankiyavashsubramanian08},
where a queueing distortion measure is considered. This is no
coincidence, since the Poisson channel is closely related to the
queueing channel introduced in \cite{anantharamverdu96}.

We also show that the Poisson process is the most difficult to cover,
in the sense that any point process that, with high probability, has no
more than $\lambda T$ points in $[0,T]$ can be described with
$-\lambda \log D$ bits per second. This is even true if an adversary
selects the point pattern provided that the pattern contains no more
than $\lambda$ points per second and that the encoder and the
reconstructor are allowed to use random codes.

Finally, we consider a Wyner-Ziv setting \cite{wynerziv76} of the
problem where some points in the pattern are known to the
reconstructor but the encoder does not know which ones they are. This
can be viewed as a dual problem to the Poisson channel with
noncausal side-information \cite{brosslapidothWLG09}. We
show that in this setting one can achieve the same minimum rate as
when the transmitter \emph{does} know the reconstructor's side-information.

The rest of this paper is arranged as follows: in
Section~\ref{sec:notations} we introduce some notation; in
Section~\ref{sec:poisson} we present the result for the Poisson process;
in Section~\ref{sec:general} we present the results for general point
processes and arbitrary point patterns; and in Section~\ref{sec:wz} we
present the results for the Wyner-Ziv setting.

\section{Notation}\label{sec:notations}
We use a lower-case letter like $x$ to denote a number, and an upper-case
letter like $X$ to denote a random variable. We use a boldface
lower-case letter like $\vect{x}$ to denote a vector, a function of
reals, or a point pattern, and it will be clear from the
context which one we mean. If $\vect{x}$ is a vector, $x_i$ denotes its
$i$th element. If $\vect{x}$ is a function, $x(t)$
denotes its value at $t\in\Reals$. If $\vect{x}$ is a point pattern,
we use $n_\vect{x}(\cdot)$ to denote its counting function, so
$n_\vect{x}(t_2)-n_\vect{x}(t_1)$ is the number of points in
$\vect{x}$ that fall in the interval $(t_1,t_2]$. We use a bold-face
upper-case letter like $\vect{X}$ to denote a random vector, a random
function, or a random point process. The random counting function
corresponding to a point process $\vect{X}$ is denoted by
$N_{\vect{X}}(\cdot)$. 

We use $\textnormal{Ber}(p)$ to denote the Bernoulli distribution of
parameter $p$, namely, the distribution that has probability $p$ on
the outcome $1$ and probability $(1-p)$ on the outcome $0$.

\section{Covering a Poisson Process}\label{sec:poisson}

Consider a homogeneous Poisson process $\mathbf{X}$ of
intensity~$\lambda$ on the interval $[0,T]$. Its counting function
$N_{\vect{X}}(\cdot)$ satisfies
\begin{equation*}
  \Pr\left[N_{\vect{X}}(t+\tau)-N_{\vect{X}}(t)=k\right] = \frac{e^{-\lambda
      \tau}(\lambda\tau)^k}{k!}
\end{equation*}
for all $\tau\in[0,T]$, $t\in[0,T-\tau]$ and $k\in\{0,1,\ldots\}$.

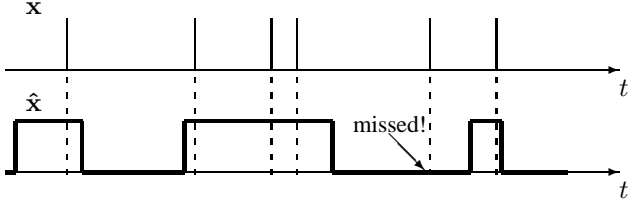
\begin{figure}[htbp]
\centering

\setlength{\unitlength}{0.68cm}
\begin{picture}(12.5,4.5)

\put(0.4,3.7){$\mathbf{x}$}
\put(0,2.6){\vector(1,0){12}}
\put(12,2.1){$t$}

\put(1.2,2.6){\line(0,1){1}}
\put(3.7,2.6){\line(0,1){1}}
\put(5.2,2.6){\line(0,1){1}}
\put(5.7,2.6){\line(0,1){1}}
\put(8.3,2.6){\line(0,1){1}}
\put(9.6,2.6){\line(0,1){1}}

\multiput(1.2,2.35)(0,-0.3){7}{\line(0,1){0.1}}
\multiput(3.7,2.35)(0,-0.3){7}{\line(0,1){0.1}}
\multiput(5.2,2.35)(0,-0.3){7}{\line(0,1){0.1}}
\multiput(5.7,2.35)(0,-0.3){7}{\line(0,1){0.1}}
\multiput(8.3,2.35)(0,-0.3){7}{\line(0,1){0.1}}
\multiput(9.6,2.35)(0,-0.3){7}{\line(0,1){0.1}}

\put(0,0.6){\vector(1,0){12}}
\put(12,0.1){$t$}

\linethickness{0.5mm}

\put(0,0.6){\line(1,0){0.2}}
\put(0.2,0.6){\line(0,1){1}}
\put(0.2,1.6){\line(1,0){1.3}}
\put(1.5,0.6){\line(0,1){1}}
\put(1.5,0.6){\line(1,0){2}}
\put(3.5,0.6){\line(0,1){1}}
\put(3.5,1.6){\line(1,0){2.9}}
\put(6.4,0.6){\line(0,1){1}}
\put(6.4,0.6){\line(1,0){2.7}}
\put(9.1,0.6){\line(0,1){1}}
\put(9.1,1.6){\line(1,0){0.6}}
\put(9.7,0.6){\line(0,1){1}}
\put(9.7,0.6){\line(1,0){1.3}}

\put(0.4,1.8){$\mathbf{\hat{x}}$}

\put(6.8,1.4){\small{missed!}}
\put(7.5,1.35){\vector(1,-1){0.7}}

\end{picture}

  \caption{Illustration of the problem.}
  \label{fig:problem-illustration}
\end{figure}

The encoder maps the realization of the Poisson process to a message
in $\{1,\ldots,2^{TR}\}$. The reconstructor then maps this message
to a $\{ 0,1 \}$-valued, Lebesgue-measurable, signal $\hat{x}(t)$, $t\in
[0,T]$. We wish to minimize the total length of the region where
$\hat{x}(t)=1$ while guaranteeing that all points in the original
Poisson process lie in this region. See
Figure~\ref{fig:problem-illustration} for an illustration.

More formally, we formulate this problem as a continuous-time
rate-distortion problem, where the distortion between the point
pattern $\mathbf{x}$ and the reproduction signal 
$\hat{\mathbf{x}}$ is 
\begin{equation}\label{eq:distortion}
  d(\mathbf{x},\hat{\mathbf{x}}) \triangleq \begin{cases}
    \frac{\mu\left(\hat{x}^{-1}(1)\right)}{T}, & \textnormal{if
      all points in $\mathbf{x}$ are in $\hat{x}^{-1}(1)$}\\
    \infty, & \textnormal{otherwise}\end{cases}
\end{equation}
where $\mu(\cdot)$ denotes the Lebesgue measure.

We say that $(R,D)$ is an achievable rate-distortion pair for the
homogeneous Poisson process of intensity $\lambda$ if, for every
$\epsilon>0$, there exists some $T_0>0$ 
such that, for every $T>T_0$, there exists an encoder
$f_T(\cdot)$ and a
reconstructor $\phi_T(\cdot)$ of rate $R+\epsilon$ bits per second
which, when applied to the Poisson 
process $\vect{X}$ on $[0,T]$, gives
\begin{equation*}
  \E{d\bigl(\mathbf{X},\phi_T\left(f_T({\mathbf{X}})\right)\bigr)} \le
  D+\epsilon. 
\end{equation*}
Denote by $R(D,\lambda)$ the minimum rate $R$ such that $(R,D)$ is
achievable for the homogeneous Poisson process of intensity
$\lambda$. Define
\begin{equation}\label{eq:poisson}
  R_{\textnormal{Pois}}(D,\lambda)\triangleq \begin{cases} -\lambda\log D  \textnormal{ bits per
        second},& D\in(0,1)\\ 0, & D\ge 1.\end{cases}
\end{equation}
\begin{thm}\label{thm:poisson}
  For all $D,\lambda>0$,
  \begin{equation}\label{eq:poisson1}
    R(D,\lambda)=R_{\textnormal{Pois}}(D,\lambda).
  \end{equation}
\end{thm}

To prove Theorem \ref{thm:poisson}, we propose a scheme to reduce the
original problem to one for a discrete memoryless source. This is
reminiscent of Wyner's
scheme for reducing the peak-limited Poisson channel to a discrete
memoryless channel \cite{wyner88}. We shall
show the optimality of this scheme in Lemma~\ref{lem:optimality}, and
we shall
then prove Theorem~\ref{thm:poisson} by computing the best rate that
is achievable using this scheme.

\emph{Scheme 1:} We divide the time-interval
$[0,T]$ into slots of $\Delta$ seconds long. The encoder first maps
the original point pattern $\vect{x}$ to a $\{0,1\}$-valued vector
$\vect{x}'$ of length $\frac{T}{\Delta}$\footnote{When $T$ is not
  divisible by $\Delta$, we consider $\vect{x}$ as a
  pattern on $[0,T']$ where $T'=\lceil \frac{T}{\Delta}\rceil
  \Delta$. When we let $\Delta$ tend to zero, the difference between
  $T$ and $T'$ also tends to zero. Henceforth we ignore this technicality
  and assume $T$
  is divisible by $\Delta$.} in the following way: if
$\vect{x}$ has at least one point in the time-slot $((i-1)\Delta,
i\Delta]$, choose $x_i'=1$; otherwise choose $x_i'=0$. The encoder
then maps $\vect{x}'$ to a message in $\{1,\ldots,2^{TR}\}$. 

Based on the encoder's message, the reconstructor produces a
$\{0,1\}$-valued length-$\frac{T}{\Delta}$ vector $\hat{\vect{x}}'$ to
meet the distortion criterion
\begin{equation*}
  \E{d'(\vect{X}',\hat{\vect{X}}')} \le D+\epsilon,
\end{equation*}
where the distortion measure $d'(\cdot,\cdot)$ is given by
\begin{IEEEeqnarray*}{rCl}
  d'(0,0)& = & 0\\
  d'(0,1)& = & 1\\
  d'(1,0)& = & \infty\\
  d'(1,1)& = & 1.
\end{IEEEeqnarray*}
It then maps $\hat{\vect{x}}'$ to a continuous-time signal
$\hat{\vect{x}}$ through
\begin{equation*}
  \hat{x}(t)=\hat{x}_{\lceil \frac{t}{\Delta} \rceil}',\quad t\in[0,T].
\end{equation*}

Scheme~1 reduces the task of designing a code for
$\vect{X}$ subject to distortion $d(\cdot,\cdot)$
to the task of designing a code for the vector
$\vect{X}'$ subject to the distortion $d'(\cdot,\cdot)$. The way we
define $d'(\cdot,\cdot)$ yields the simple relation
\begin{equation}
  d(\vect{x},\hat{\vect{x}})=d'(\vect{x}',\hat{\vect{x}}').
\end{equation}
When $\vect{X}$ is the homogeneous Poisson process of intensity
$\lambda$, the components of $\vect{X}'$ are
independent and identically distributed (IID) 
$\textnormal{Ber}(1-e^{-\lambda \Delta})$. Let $R_\Delta(D,\lambda)$
denote the rate-distortion function for $\vect{X}'$ and
$d'(\cdot,\cdot)$. If we combine
Scheme~1 with an optimal code for $\vect{X}'$ subject to
$\E{d'(\vect{X}', \hat{\vect{X}}')} < D+\epsilon$, we can achieve any
rate that is larger than
\begin{equation*}
  \frac{R_\Delta(D,\lambda) \textnormal{ bits}}{\Delta \textnormal{
      seconds}}.
\end{equation*}

The next lemma, which is reminiscent of
\cite[Theorem 2.1]{wyner88b}, shows that when we let $\Delta$ tend to
zero, there is no loss in optimality in using Scheme~1.

\begin{lem}\label{lem:optimality}
  For all $D,\lambda > 0$,
  \begin{equation}\label{eq:lem}
    R(D,\lambda)=\lim_{\Delta\downarrow 0}
    \frac{R_\Delta(D,\lambda)}{\Delta}.
  \end{equation}
\end{lem}
\begin{IEEEproof} See Appendix.
\end{IEEEproof}
  
\begin{IEEEproof}[Proof of Theorem \ref{thm:poisson}]
We derive $R(D,\lambda)$ by computing the  
right-hand side of \eqref{eq:lem}. To compute $R_\Delta(D,\lambda)$ we
apply Shannon's formula of the rate-distortion function for a discrete
memoryless source \cite{shannon48}:
\begin{equation}\label{eq:1}
  R_\Delta(D,\lambda)=\min_{P_{\hat{Z}|Z}: \E{d_\Delta(Z,\hat{Z}})\le D}
  I(Z;\hat{Z}). \footnote{Strictly speaking, since our distortion
    measure is unbounded, we need to modify Shannon's proof of this
    formula in order to use it for our problem. This can be done
    by letting the reconstructor produce the 
    all-one sequence, which yields bounded distortion for any source
    sequence, whenever no codeword can be found that is jointly typical
    with the source sequence.}
\end{equation}
When $D\in(0,1)$, the
conditional distribution $P_{\hat{Z}|Z}$ which achieves the minimum on
the right-hand side of \eqref{eq:1} is
\begin{IEEEeqnarray*}{rCl}
  P_{\hat{Z}|Z}^*(1|0) & = & D e^{\lambda\Delta}-e^{\lambda\Delta}+1,\\
  P_{\hat{Z}|Z}^*(1|1) & = & 1.
\end{IEEEeqnarray*}
Computing the mutual information $I(Z;\hat{Z})$ under this
$P_{\hat{Z}|Z}^*$ yields
\begin{equation}\label{eq:2}
  R_\Delta(D,\lambda)=\Hb(D)-e^{-\lambda\Delta}\Hb(D
    e^{\lambda\Delta}- e^{\lambda\Delta}+1),\ \  D\in(0,1),
\end{equation}
where $\Hb(\cdot)$ denotes the binary entropy function.

When $D\ge 1$, it is optimal to choose $\hat{Z}=1$
(deterministically), yielding 
\begin{equation}\label{eq:3}
  R_\Delta(D,\lambda)=0,\quad D\ge 1.
\end{equation}

Combining \eqref{eq:lem}, \eqref{eq:2} and \eqref{eq:3} and computing
the limit as $\Delta$ tends to zero yields
\eqref{eq:poisson1}.  
\end{IEEEproof}

\section{Covering General Point Processes and Arbitrary Point
  Patterns}\label{sec:general}  
We next consider a general point process $\mathbf{Y}$. 
We assume that there exists some $\lambda$ such that
\begin{equation}\label{eq:general}
  \lim_{t\to\infty} \Pr \left[
    \frac{N_{\vect{Y}}(t)}{t}>\lambda+\delta\right] = 0 \quad \textnormal{for
    all } \delta>0. 
\end{equation}
Condition \eqref{eq:general} is satisfied, for example, when
$\mathbf{Y}$ is an ergodic process whose expected number of points per
second is less than or equal to $\lambda$. 

Since the Poisson process is memoryless, one naturally expects it to
be the most difficult to describe. This is indeed the case, as the
next theorem shows.

\begin{thm}\label{thm:general}
  The pair $(R_{\textnormal{Pois}}(D,\lambda),D)$ is
  achievable on any point process 
  satisfying~\eqref{eq:general}. 
\end{thm}

Before proving Theorem \ref{thm:general}, we state a stronger
result. Consider a point pattern $\vect{z}$ chosen by an adversary
on the interval $[0,T]$ which contains no more than $\lambda T$ 
points. The corresponding counting function
$n_{\vect{z}}(\cdot)$ must then satisfy
\begin{equation}\label{eq:arbitrary}
  n_{\vect{z}}(T) \le \lambda T.
\end{equation}
The encoder and the reconstructor are
allowed to use random codes. Namely, they fix a distribution on all
(deterministic) codes of a certain rate on $[0,T]$. According to this
distribution, they randomly pick a code which is not revealed to the
adversary. They then apply it to the point pattern $\vect{z}$ chosen
by the adversary. We say that
$(R,D)$ is achievable with random coding against an adversary subject
to \eqref{eq:arbitrary} if, for every $\epsilon>0$, there exists some
$T_0$ such that, for every $T>T_0$, there exists a random code
on $[0,T]$ of rate $R+\epsilon$ such that the expected
distortion between \emph{any} $\vect{z}$ satisfying
\eqref{eq:arbitrary} and its reconstruction is smaller than $D+\epsilon$.

\begin{thm}\label{thm:arbitrary}
  The pair $(R_{\textnormal{Pois}}(D,\lambda),D)$ is
  achievable with random coding against an adversary subject
  to~\eqref{eq:arbitrary}. 
\end{thm}
\begin{IEEEproof}
  First note that when $D\ge 1$, the encoder does not need to describe
  the pattern: the reconstructor simply produces the all-one
  function, yielding distortion $1$ for any $\vect{z}$. Hence the pair
  $(0,D)$ is achievable with random coding.
  
  Next consider $D\in(0,1)$. We use Scheme~1 as in Section
  \ref{sec:poisson} to reduce the original problem to one of random
  coding for an arbitrary discrete-time sequence $\vect{z}'$. Here
  $\vect{z}'$ is $\{0,1\}$-valued, has length $\frac{T}{\Delta}$, and
  satisfies 
  \begin{equation}\label{eq:constraint_discrete}
    \sum_{i=1}^{T/\Delta} z_i' \le \lambda T.
  \end{equation}
  We shall construct a random code of rate $\frac{R}{\Delta}$ which,
  when applied to any $\vect{z}'$ satisfying
  \eqref{eq:constraint_discrete}, yields
  \begin{equation*}
    \E{d'(\vect{z}',\hat{\vect{Z}}')} < D+\epsilon,
  \end{equation*}
  where the random vector $\hat{\vect{Z}}'$ is the result of applying
  the random encoder and decoder to $\vect{z}'$. Combined with
  Scheme~1 this random code will yield a random code on the
  continuous-time point pattern $\vect{z}$ that achieves the
  rate-distortion pair $(R,D)$.
  
  Our discrete-time random code consists of $2^{TR}$ $\{0,1\}$-valued,
  length-$\frac{T}{\Delta}$ random sequences $\hat{\vect{Z}}_m'$,
  $m\in\{1,\ldots, 2^{TR}\}$. The first sequence $\hat{\vect{Z}}_1'$
  is chosen deterministically to be the all-one sequence. The other
  $2^{TR}-1$ sequences are drawn independently, with each sequence
  drawn IID $\textnormal{Ber}(D)$.
  
  To describe source sequence $\vect{z}'$, the encoder looks for a
  codeword $\hat{\vect{z}}_m'$, $m\in\{2,\ldots,2^{TR}\}$  such that
  \begin{equation}\label{eq:11}
    \hat{z}_{m,i}'=1 \textnormal{ whenever }z_i'=1.
  \end{equation}
  If it finds one or more such codewords, it sends the index of the
  first one; otherwise it sends $1$. The
  reconstructor outputs the 
  sequence $\hat{\vect{z}}_m'$ where $m$ is the message it receives
  from the encoder.
  
  We next analyze the expected distortion of this random code for a
  fixed $\vect{z}'$ satisfying \eqref{eq:constraint_discrete}. Define
  \begin{equation*}
    \mu\triangleq \frac{\sum_{i=1}^{T/\Delta} z_i'}{T},
  \end{equation*}
  and note that by \eqref{eq:constraint_discrete} $\mu\le
  \lambda$. Denote by $\mathcal{E}$ the 
  event that the encoder cannot find $\hat{\vect{z}}_m'$,
  $m\in\{2,\ldots, 2^{TR}\}$ satisfying \eqref{eq:11}. If
  $\mathcal{E}$ occurs, the encoder sends $1$ and the resulting
  distortion is equal to $1$. 
  
  The probability that a randomly
  drawn codeword $\hat{\vect{Z}}_m'$ satisfies \eqref{eq:11} is
  \begin{equation*}
    D^{\mu T}\ge D^{\lambda T} = 2^{(\lambda \log D)T}.
  \end{equation*}
  Because the codewords $\hat{\vect{Z}}_m'$, $m\in\{2,\ldots,2^{TR}\}$
  are chosen independently, if we choose $R>-\lambda \log D$, then
  $\Pr [\mathcal{E}] \to 0$ as $T\to\infty$. Hence, for large
  enough $T$, the contribution to the expected distortion from the
  event $\mathcal{E}$ can be ignored.
  
  We next analyze the expected distortion conditional on
  $\mathcal{E}^{\textnormal{c}}$. The reproduction $\hat{\vect{Z}}'$ has the
  following distribution: at positions where $\vect{z}'$ takes the value
  $1$, $\hat{\vect{Z}}'$ must also be $1$; at other positions the
  elements of $\hat{\vect{Z}}'$ have the IID $\textnormal{Ber}(D)$
  distribution. Thus the expected value of $\sum_{i=1}^{T/\Delta}
  \hat{Z}_i'$ is 
  $\mu T + D(\frac{T}{\Delta}-\mu T)$, and
  \begin{equation*}
    \E{\left.d'(\vect{z}',\hat{\vect{Z}}')\right| \mathcal{E}^\textnormal{c}}
    = D+ (1-D)\mu\Delta.
  \end{equation*}
  When we let $\Delta$ tend to zero, this value tends to
  $D$. We have thus shown that, for small enough $\Delta$, we can
  achieve the pair $(R/\Delta, D)$ on $\vect{z}'$ with random coding
  whenever $R>-\lambda \log D$, and therefore we can also achieve
  $(R,D)$ on the continuous-time point pattern $\vect{z}$ with random
  coding if $R>-\lambda\log D$. 
\end{IEEEproof}

We next use Theorem~\ref{thm:arbitrary} to prove
Theorem~\ref{thm:general}.

\begin{IEEEproof}[Proof of Theorem~\ref{thm:general}]  
It follows from
Theorem~\ref{thm:arbitrary} that, on any point process satisfying
\eqref{eq:general}, the pair $(R_{\textnormal{Pois}}(D,\lambda+\delta),D)$ is achievable
with \emph{random coding}. Further, since there is no adversary, the
existence of a good random code guarantees the existence of
a good deterministic code. Hence $(R_{\textnormal{Pois}}(D,\lambda+\delta),D)$ is also
achievable on this process with deterministic
coding. Theorem~\ref{thm:general} now follows when we let $\delta$
tend to zero, since $R_{\textnormal{Pois}}(D,\cdot)$ is a continuous function.
\end{IEEEproof}

\section{Some Points are Known to  the  Reconstructor}\label{sec:wz} 

In this section we consider a Wyner-Ziv setting for our
problem. We first consider the case where $\vect{X}$ is a homogeneous
Poisson process of intensity $\lambda$. (Later we consider an
arbitrary point pattern.) Assume that each point in $\vect{X}$ is
known to the reconstructor independently with  
probability $p$. Also assume that the encoder does not know which
points are known to the reconstructor. The encoder maps $\vect{X}$
to a message in $\{1,\ldots,2^{TR}\}$, and the reconstructor 
produces a Lebesgue-measurable, $\{0,1\}$-valued signal $\hat{\vect{X}}$ on
$[0,T]$ based on this message and the positions of the points that he
knows. The achievability of a rate-distortion pair is defined
in the same way as in  
Section~\ref{sec:poisson}. Denote the smallest rate $R$ for which
$(R,D)$ is achievable by $R_{\textnormal{WZ}}(D,\lambda,p)$.

Obviously, $R_{\textnormal{WZ}}(D,\lambda,p)$ is lower-bounded by the
smallest achievable rate when the transmitter \emph{does} know which
points are known to the reconstructor. The latter rate is given by
$R_{\textnormal{Pois}}(D,(1-p)\lambda)$, where
$R_{\textnormal{Pois}}(\cdot,\cdot)$ is given by 
\eqref{eq:poisson}. Indeed, when the encoder knows which points are 
known to the reconstructor, it is optimal for it to describe only the
remaining points, which themselves form a homogeneous Poisson process
of intensity $(1-p)\lambda$. The reconstructor then selects a
set based on this description to cover the points unknown to it and
adds to this set the points it knows. Thus,
\begin{equation}\label{eq:wz1}
  R_{\textnormal{WZ}}(D,\lambda,p)\ge R_{\textnormal{Pois}}(D,(1-p)\lambda).
\end{equation}
The next theorem shows that \eqref{eq:wz1} holds with equality. 

\begin{thm}\label{thm:wz}
  Knowing the points at the reconstructor only is as good as knowing
  them also at the encoder:
  \begin{equation}
    R_{\textnormal{WZ}}(D,\lambda,p) = R_{\textnormal{Pois}}(D,(1-p)\lambda).
  \end{equation}
\end{thm}

To prove Theorem~\ref{thm:wz}, it remains to show that the pair
$(R_{\textnormal{Pois}}(D,(1-p)\lambda), D)$ is achievable. We shall
show this as a 
consequence of a stronger result concerning arbitrarily varying
sources.

Consider an arbitrary point pattern $\vect{z}$ on $[0,T]$ chosen by an
adversary. The adversary is allowed to put at most $\lambda T$ points
in $\vect{z}$. Also, it must reveal all but at most $\nu T$
points to the reconstructor, without telling the encoder which points
it has revealed. The encoder and the reconstructor are
allowed to use random codes, where the encoder is a random mapping
from $\vect{z}$ to a message in $\{1,\ldots, 2^{TR}\}$, and where the
reconstructor is a random mapping from this message, together with the point
pattern that it knows, to a $\{0,1\}$-valued, Lebesgue-measurable
signal $\hat{\vect{z}}$. The distortion $d(\vect{z},\hat{\vect{z}})$
is defined as in \eqref{eq:distortion}. 

\begin{thm}\label{thm:wzadversary}
  Against an adversary who puts at most $\lambda T$ points on $[0,T]$
  and reveals all but at most $\nu T$ points to the reconstructor, the
  rate-distortion pair $(R_{\textnormal{Pois}}(D,\nu),D)$ is
  achievable with random coding. 
\end{thm}

\begin{IEEEproof}
  The case $D\ge 1$ is trivial, so we shall only consider the
  case where $D\in(0,1)$. The encoder 
  and the reconstructor first use Scheme~1 as in 
  Section~\ref{sec:poisson} to reduce the point pattern $\vect{z}$ to a
  $\{0,1\}$-valued vector $\vect{z}'$ of length
  $\frac{T}{\Delta}$. Define
  \begin{equation*}
    \mu\triangleq \frac{\sum_{i=1}^{T/\Delta} z_i'}{T},
  \end{equation*}
  and note that, by assumption, $\mu\le\lambda$. If $\mu\le \nu$, then
  we can ignore the reconstructor's side-information and use the 
  random code of Theorem~\ref{thm:arbitrary}. Henceforth we assume
  $\mu>\nu$. 
  
  Denote by $\vect{s}$ the point pattern known to the reconstructor
  and by $\vect{s}'$ the vector obtained from $\vect{s}$ through the
  discretization in time of Scheme~1. Since there are at most $\nu T$
  points that are unknown to the reconstructor, 
  \begin{equation}\label{eq:14}
    \sum_{i=1}^{T/\Delta} s_i'\ge (\mu-\nu)T.
  \end{equation}
  
  The encoder conveys the value of $\mu T$ to the receiver using
  bits. Since $\mu T$ is an integer between $0$ and $\lambda T$, the
  number of 
  bits per second needed to describe it tends to zero as $T$ tends to
  infinity.
  
  Next, the encoder and the reconstructor randomly generate
  $2^{T(R+\tilde{R})}$ independent codewords $$\hat{\vect{z}}_{m,l}',\quad 
  m\in\{1,\ldots, 2^{TR}\},\ l\in\{1,\ldots,2^{T\tilde{R}}\},$$
  where each codeword is generated IID $\textnormal{Ber}(D)$.
  
  To describe $\vect{z}'$, the encoder looks for a codeword
  $\hat{\vect{z}}_{m,l}'$ such that
  \begin{equation}\label{eq:12}
    \hat{z}_{m,l,i}'=1 \textnormal{ whenever } z_i'=1.
  \end{equation}
  If it finds one or more such codewords, it sends the index $m$ of
  the first one; otherwise
  it tells the reconstructor to produce the all-one sequence.
  
  When the reconstructor receives the index $m$, it looks for an index
  $\tilde{l}\in\{1,\ldots,2^{T\tilde{R}}\}$ such that
  \begin{equation}\label{eq:13}
    \hat{z}_{m,\tilde{l},i}'=1 \textnormal{ whenever }
    s_i'=1.
  \end{equation}
  If there is only one such codeword, it outputs it as the
  reconstruction; if there are more than one such codewords, it
  outputs the all-one sequence.  
  
  To analyze the expected distortion for $\vect{z}'$ over this random
  code, first consider the event that the encoder cannot find a
  codeword satisfying \eqref{eq:12}. Note that the probability that a
  randomly generated codeword satisfies \eqref{eq:12} is $D^{\mu
    T}$, so the probability of this event tends to zero as
  $T$ tends to infinity provided that
  \begin{equation}\label{eq:15}
    R+\tilde{R}>-\mu \log D.
  \end{equation}
  
  Next consider the event that the reconstructor finds more than one
  $\tilde{l}$ satisfying \eqref{eq:13}. 
  The probability that a randomly generated codeword satisfies
  \eqref{eq:13} is $D^{\sum_{i=1}^{T/\Delta} s_i'}$. Consequently, by
  \eqref{eq:14} the probability of this event tends to zero as
  $T$ tends to infinity provided
  \begin{equation}\label{eq:16}
    \tilde{R} < -(\mu-\nu)\log D.
  \end{equation}
  
  Finally, if the encoder finds a codeword satisfying \eqref{eq:12}
  and the reconstructor finds only one codeword satisfying
  \eqref{eq:13}, then the two codewords must be the same. Following the
  same calculations as in the proof of Theorem~\ref{thm:arbitrary},
  the expected distortion in this case tends to $D$ as $\Delta$ tends
  to zero.
  
  Combining \eqref{eq:15} and \eqref{eq:16}, we can make the expected
  distortion arbitrarily close to $D$ as $T\to\infty$ if
  \begin{equation*}
    R>-\nu \log D.
  \end{equation*}
\end{IEEEproof}

\begin{IEEEproof}[Proof of Theorem \ref{thm:wz}]
  The claim follows from \eqref{eq:wz1}, Theorem
  \ref{thm:wzadversary}, and the Law of Large Numbers.
\end{IEEEproof}

\begin{appendix}
  In this appendix we prove Lemma~\ref{lem:optimality}. Given
  any rate-distortion code with $2^{TR}$ codewords
  $\hat{\vect{x}}_m$, $m\in\{1,\ldots, 2^{TR}\}$ that achieves
  expected distortion $D$, we shall construct a new code that can be
  constructed through Scheme~1, that contains $(2^{TR}+1)$ codewords, and
  that achieves an expected distortion that is arbitrarily close to $D$.
  
\begin{figure}[htbp]
\centering

\setlength{\unitlength}{0.68cm}
\begin{picture}(12,4.5)

\put(0,3){\vector(1,0){12}}
\put(12,2.5){$t$}
\multiput(0,2.9)(1,0){12}
{\line(0,1){0.2}}
\multiput(0,2.7)(1,0){12}
{\line(0,1){0.1}}
\multiput(0,2.4)(1,0){12}
{\line(0,1){0.1}}
\multiput(0,2.1)(1,0){12}
{\line(0,1){0.1}}
\multiput(0,1.8)(1,0){12}
{\line(0,1){0.1}}
\multiput(0,1.5)(1,0){12}
{\line(0,1){0.1}}
\multiput(0,1.2)(1,0){12}
{\line(0,1){0.1}}




\put(2.35,2.65){\tiny{$\Delta$}}
\put(2.3,2.75){\vector(-1,0){0.3}}
\put(2.7,2.75){\vector(1,0){0.3}}





\put(0,1){\vector(1,0){12}}
\put(12,0.5){$t$}
\multiput(0,0.9)(1,0){12}
{\line(0,1){0.2}}

\linethickness{0.5mm}
\put(0,3){\line(1,0){0.2}}
\put(0.2,3){\line(0,1){1}}
\put(0.2,4){\line(1,0){1.3}}
\put(1.5,3){\line(0,1){1}}
\put(1.5,3){\line(1,0){2}}
\put(3.5,3){\line(0,1){1}}
\put(3.5,4){\line(1,0){2.9}}
\put(6.4,3){\line(0,1){1}}
\put(6.4,3){\line(1,0){2.7}}
\put(9.1,3){\line(0,1){1}}
\put(9.1,4){\line(1,0){0.6}}
\put(9.7,3){\line(0,1){1}}
\put(9.7,3){\line(1,0){1.3}}

\put(0.2,4.2){$\mathbf{1}_{\mathcal{A}_m}$}

\put(0,1){\line(0,1){1}}
\put(0,2){\line(1,0){2}}
\put(2,1){\line(0,1){1}}
\put(2,1){\line(1,0){1}}
\put(3,1){\line(0,1){1}}
\put(3,2){\line(1,0){4}}
\put(7,1){\line(0,1){1}}
\put(7,1){\line(1,0){2}}
\put(9,1){\line(0,1){1}}
\put(9,2){\line(1,0){1}}
\put(10,1){\line(0,1){1}}
\put(10,1){\line(1,0){1}}

\put(0.2,2.2){$\mathbf{\hat{w}}_m$}

\end{picture}

  \caption{Constructing $\hat{\vect{w}}_m$ from
    $\mathcal{A}_m$.}
  \label{fig:discretize}
\end{figure}
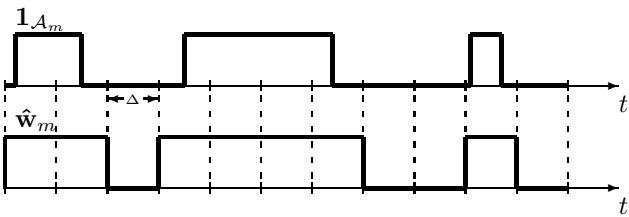

  Denote the codewords of our new code by $\hat{\vect{w}}_m$,
  $m\in\{1,\ldots,2^{TR}+1\}$. We choose the last codeword to be the
  constant 1. We next describe our choices for the other codewords.
  For every $\epsilon>0$ and every $\hat{\vect{x}}_m$, we can
  approximate the set $\{t\colon \hat{x}_m(t)=1\}$ by a set
  $\mathcal{A}_m$  that is equal to a finite, say $N_m$, union of open
  intervals. More specifically,
  \begin{equation}\label{eq:21}
    \mu\left(\hat{x}_m^{-1}(1)\bigtriangleup
      \mathcal{A}_m\right)\le 2^{-TR}\epsilon,
  \end{equation}
  where $\bigtriangleup$ denotes the
  symmetric difference between two sets (see,
  e.g., \cite[Chapter 3, Proposition 15]{royden88}). Define
  \begin{equation*}
    \set{B}\triangleq \bigcup_{m=1}^{2^{TR}}
    \left(\hat{x}_m^{-1}(1)\setminus \mathcal{A}_m\right),
  \end{equation*}
  and note that by \eqref{eq:21}
  \begin{equation}\label{eq:19}
    \mu (\set{B})\le \epsilon.
  \end{equation}
  For each $\mathcal{A}_m$, $m\in\{1,\ldots,2^{TR}\}$, define
  \begin{equation*}
    \mathcal{T}_m\triangleq \left\{t\in [0,T]\colon \bigl( \left(\left\lceil
            {t}/{\Delta}\right\rceil - 
          1\right)\Delta, \left\lceil
          {t}/{\Delta}\right\rceil\Delta\bigr] \cap 
        \mathcal{A}_m \neq\emptyset\right\}. 
  \end{equation*}
  We now construct $\hat{\vect{w}}_m$, $m\in\{1,\ldots,2^{TR}\}$ as
  \begin{equation*}
    \hat{\vect{w}}_m= \mathbf{1}_{\mathcal{T}_m},
  \end{equation*}
  where $\mathbf{1}_\mathcal{S}$ denotes the indicator function of the
  set $\mathcal{S}$. Note that $\mathcal{A}_m \subseteq \mathcal{T}_m
  = \hat{w}_m^{-1}(1)$.
  See Figure~\ref{fig:discretize} for an illustration of this construction. 
Let
 $$ N\triangleq \max_{m\in\{1,\ldots,2^{TR}\}} N_m.$$ 
 It can be seen that 
  \begin{equation}\label{eq:20}
    \mu\left(\hat{w}_m^{-1}(1)\right)-\mu(\mathcal{A}_m) \le
  2N\Delta, \quad m\in\{1,\ldots,2^{TR}\}.
  \end{equation}
  Our encoder works as follows: if $\vect{x}$ contains no point in
  $\mathcal{B}$, it maps $\vect{x}$ to the same message as the given
  encoder; otherwise it maps $\vect{x}$ to the index $(2^{TR}+1)$ of
  the all-one codeword. To analyze the distortion, first consider the
  case where $\vect{x}$ contains no point in $\mathcal{B}$. In this 
  case, all points in $\vect{x}$ must be covered by the selected codeword
  $\hat{\vect{w}}_m$. By \eqref{eq:21} and \eqref{eq:20}, the
  difference
  $d(\vect{x},\hat{\vect{w}}_m)-d(\vect{x},\hat{\vect{x}}_m)$, if
  positive, can be  
  made arbitrarily small by choosing small $\epsilon$ and
  $\Delta$. Next consider the case where $\vect{x}$ does contain
  points in $\mathcal{B}$. By \eqref{eq:19}, the probability that this
  happens can be made arbitrarily small by choosing $\epsilon$ small,
  therefore its contribution to the expected distortion can also be made
  arbitrarily small. We conclude that our code
  $\{\hat{\vect{w}}_m\}$ can achieve a distortion that is arbitrarily
  close to the distortion achieved by the original code
  $\{\hat{\vect{x}}_m\}$. This concludes the proof of
  Lemma~\ref{lem:optimality}. 
\end{appendix}

\bibliographystyle{IEEEtran}           
\bibliography{/Volumes/Data/wang/Library/texmf/tex/bibtex/header_short,/Volumes/Data/wang/Library/texmf/tex/bibtex/bibliofile}

\begin{thebibliography}{10}
\providecommand{\url}[1]{#1}
\def\UrlFont{\rmfamily}
\providecommand{\newblock}{\relax}
\providecommand{\bibinfo}[2]{#2}
\providecommand\BIBentrySTDinterwordspacing{\spaceskip=0pt\relax}
\providecommand\BIBentryALTinterwordstretchfactor{4}
\providecommand\BIBentryALTinterwordspacing{\spaceskip=\fontdimen2\font plus
\BIBentryALTinterwordstretchfactor\fontdimen3\font minus
  \fontdimen4\font\relax}
\providecommand\BIBforeignlanguage[2]{{%
\expandafter\ifx\csname l@#1\endcsname\relax
\typeout{** WARNING: IEEEtran.bst: No hyphenation pattern has been}%
\typeout{** loaded for the language `#1'. Using the pattern for}%
\typeout{** the default language instead.}%
\else
\language=\csname l@#1\endcsname
\fi
#2}}

\bibitem{kabanov78}
Y.~Kabanov, ``The capacity of a channel of the {P}oisson type,'' \emph{Theory
  of Probability and Its Appl.}, vol.~23, pp. 143--147, 1978.

\bibitem{davis80}
M.~H.~A. Davis, ``Capacity and cutoff rate for {P}oisson-type channels,''
  \emph{IEEE Trans. Inform. Theory}, vol.~26, pp. 710--715, Nov. 1980.

\bibitem{wyner88}
A.~D. Wyner, ``Capacity and error exponent for the direct detection photon
  channel --- part {I},'' \emph{IEEE Trans. Inform. Theory}, vol.~34, no.~6,
  pp. 1449--1461, Nov. 1988.

\bibitem{wyner88b}
------, ``Capacity and error exponent for the direct detection photon channel
  --- part {II},'' \emph{IEEE Trans. Inform. Theory}, vol.~34, pp. 1462--1471,
  Nov. 1988.

\bibitem{coverchiang02}
T.~M. Cover and M.~Chiang, ``Duality between channel capacity and rate
  distortion with two-sided state information,'' \emph{IEEE Trans. Inform.
  Theory}, vol.~48, no.~6, pp. 1629--1638, June 2002.

\bibitem{rubin74}
I.~Rubin, ``Information rates and data-compression schemes for {P}oisson
  processes,'' \emph{IEEE Trans. Inform. Theory}, vol.~20, no.~2, pp. 200--210,
  Mar. 1974.

\bibitem{colemankiyavashsubramanian08}
T.~P. Coleman, N.~Kiyavash, and V.~G. Subramanian, ``The rate-distortion
  function of a {P}oisson process with a queueing distortion measure,'' in
  \emph{Proceedings Data Compression Conference 2008}, Cliff Lodge, Snowbird,
  Utah, USA, Mar. 2008.

\bibitem{anantharamverdu96}
V.~Anantharam and S.~Verd\'u, ``Bits through queues,'' \emph{IEEE Trans.
  Inform. Theory}, vol.~42, no.~1, pp. 4--18, Jan. 1996.

\bibitem{wynerziv76}
A.~D. Wyner and J.~Ziv, ``The rate-distortion function for source coding with
  side information at the decoder,'' \emph{IEEE Trans. Inform. Theory},
  vol.~22, no.~1, pp. 1--10, Jan. 1976.

\bibitem{brosslapidothWLG09}
S.~Bross, A.~Lapidoth, and L.~Wang, ``The {P}oisson channel with side
  information,'' in \emph{Proceedings Forty-Seventh Allerton Conf. Comm.,
  Contr. and Comp.}, Allerton House, Monticello, Illinois, September
  30--October 2, 2009.

\bibitem{shannon48}
C.~E. Shannon, ``A mathematical theory of communication,'' \emph{Bell System
  Techn. J.}, vol.~27, pp. 379--423 and 623--656, July and Oct. 1948.

\bibitem{royden88}
H.~L. Royden, \emph{Real Analysis}, 3rd~ed.\hskip 1em plus 0.5em minus
  0.4em\relax Macmillan Publishing Company, 1988.

\end{thebibliography}

\end{document}